\documentclass{cargese}\usepackage{graphicx}

\newif\ifpreprint
\preprintfalse

\let\footnote\savefootnote
\let\footnotetext\savefootnotetext
\let\footnoterule\savefootnoterule
\normallatexbib

\def\bea{\begin{eqnarray}}
\def\eea{\end{eqnarray}}
\def\beann{\begin{eqnarray*}}
\def\eeann{\end{eqnarray*}}
\def\beq{\begin{equation}}
\def\eeq{\end{equation}}
\def\ba{\begin{array}}
\def\ea{\end{array}}
\def\ben{\begin{enumerate}}
\def\een{\end{enumerate}}
\def\4{\tilde }
\def\5{\bar }
\def\6{\partial }
\def\7{\hat }
\def\ep{\epsilon}

\def\pd{\partial_}

\font\mybb=msbm10 at 10pt
\def\bb#1{\hbox{\mybb#1}}

\def\bE {\bb{E}}

\def\m{\mu}

\def\l{\lambda}

\def\x{\chi}

\def\f{\phi}
\def\d{\delta}

\begin{document}


\articletitle[]{Generalized Calibrations}


\author{Jan Gutowski}


\affil{DAMTP, Silver Street, University of Cambridge, Cambridge
CB3 9EW}

%
\email{jbg21@damtp.cam.ac.uk}
%

\begin{abstract}

We present a generalization of calibrations in which the
calibration form is not closed. We use this to examine a class of
supersymmetric p-brane worldvolume solitons.As an example we consider 
M5-brane worldvolume solitons in an AdS background.

\end{abstract}


\paragraph{Introduction}
\index{introduction}

There has been considerable progress recently in the
classification of p-brane worldvolume solitons. In a flat
background with no Born-Infeld type fields the worldvolume
dynamics are governed by the Nambu-Goto action
\beq
S_{NG}= \int d^{p+1}x \sqrt{- \det {g}}
\label{e1}
\eeq
where $g$ denotes the pull-back of the background metric to the
worldvolume. For such configurations this Lagrangian is equal to
the energy density of the p-brane and solutions to the equations
of motion minimize both the volume and the energy.

A large number of
solutions have been classified in terms of calibrated geometries
\cite{HL},\cite{CALa},\cite{CALb},\cite{CALc}. The solutions are described
by calibration forms defined  on the target space, and the brane
worldvolumes correspond to calibrated sub-manifolds of the target
space. Many examples have been constructed, and the types of
geometry encountered are typically K\"ahler, special Lagrangian and
exceptional. The resulting solitons may be seen to  arise from
multiple intersections of p-branes which preserve some proportion
of the supersymmetry depending on the geometry of the contact set.
However there are limitations to the methods used here,
they do not allow treatment of configurations with non-vanishing
Born-Infeld fields and they do not describe solitons in curved
backgrounds whose worldvolume actions are modified by the presence
of Wess-Zumino terms.

In this article we present a generalization
of the concept of a calibrated geometry which enables us to
describe a class of solitons in a curved background but with
vanishing Born-Infeld fields. It is necessary to modify the
definition of the calibration form as in a curved background the
energy is not equal to the p-brane volume.

\paragraph{Generalized Calibrations}
\index{generalized calibrations}

Let $(M,g)$ be an n-dimensional Riemannian manifold with metric
 $g$, and  $x \in M$. Suppose that $G(p,T_x (M))$ is the
  Grassmannian of (oriented) p-planes in $T_x (M)$.
 Then for $\x \in G(p,T_x (M))$, there exists an orthonormal
 basis with respect to g, $\lbrace e_1 ... e_n \rbrace$ of
$T_x (M)$ such that $\lbrace e_1 ... e_p \rbrace$ is a basis
of $\x$. The co-volume of $\x$ is
then defined as
\beq
{\buildrel \rightarrow \over \x} = e_1 \wedge ... \wedge e_p.
\label{e2}
\eeq
For our definition of a generalized calibration (referred to from
now on as simply a calibration), we drop the standard requirement
of closure for the calibration form.

A calibration of degree p on an open subset $U \subset T_x (M)$ is
a p-form $\f$ such that, at each $x \in  U$, ${\f_x} ( {\buildrel
\rightarrow \over \x}) \leq 1$ for all $\x \in G(p,T_x (M))$. It is
also required that the contact set $G( \f)$ should be
non-empty, where
\beq
G( \f )= \lbrace \x \in G(p,T_x (M)):{\f} ( {\buildrel
\rightarrow \over \x}) \leq 1 \rbrace.
\label{e3}
\eeq
Suppose now that $N$ is a p-dimensional submanifold of $M$.
Then $N$ is a calibrated submanifold (or calibration for
short) of
degree p if
\beq
\f_x ( {\buildrel \rightarrow \over N_x} ) =1
\label{e4}
\eeq
for all $x \in N$,where $\f$ is a calibration of degree p in
$T(M)$. ${\buildrel \rightarrow \over N_x}$ is the co-volume of the
tangent space $T_x N$.

As we have removed the requirement that $d \f=0$ should
 hold, it is not the case generally that $N$ is volume minimizing.
However another quantity is minimized. Suppose N is  calibrated
and U is an open sub-manifold of N and V is an open sub-manifold
of M such that $\partial U =
\partial V$. Then let $L$ be a manifold with oriented boundary
$\partial L = U -V$. We may then write
\beq
\int \m_U = \int \f ({\buildrel \rightarrow \over U})
\m_U  = \int_U \f = \int_V \f +\int_L d \f \leq \int \m_V + \int_L
d \f
\label{e5}
\eeq
where we have used Stokes' theorem and $\m$ denotes the volume
form. So the new minimized quantity
is
\beq
 \int d^p x \sqrt{\det (g)} - \int_B d \f
\label{e6}
\eeq
where B is a manifold whose boundary is N and $g$ is
 the pull-back of the metric to N. This quantity is of
considerable interest in theories for which the Born-Infeld fields
vanish, and there is a non-trivial Wess-Zumino term; such as
p-branes in a curved supergravity background, which we consider in
the next section. For these theories, N is identified with some
spatial submanifold of the worldvolume and $\partial B =N$.

\paragraph{{Examples}}
\index{examples}

We consider worldvolume solitons on a M5-brane in the background
of a stack of parallel M5-branes in the near horizon limit, so the
background metric and 4-form are given by
\bea
ds^2 = {r \over R} ds^2 \big( \bE^{5,1} \big) +{R^2 \over r^2}
\big(dr^2 +r^2 ds^2 \big( S^4 \big) \big) \\
G_4 = \m_{S^4}.
\label{e7}
\eea
$R$ is a positive constant and the above geometry is $AdS_7
\times S^4$.
We work in the static gauge and consider solutions which depend
only on the $5-q$ worldvolume co-ordinates $ \lbrace x^i: i=1,
\dots , 5-q \rbrace$. The worldvolume action may then be written
as
\beq
S = \l \left( \int d^{5-q}x \sqrt{ \det ({\tilde g_{ij}}) } - \int
{\tilde{F}} \right)
\label{e8}
\eeq
for constant $\l$ and
\bea
{\tilde{g}}_{ij}= \left({r \over R} \right)^{{q+1} \over {5-q}}
\left( {r \over R} \d_{ij} +{R^2 \over r^2}\d_{a b} \pd{i}y^a
\pd{j}y^b \right).
\label{e9}
\eea
The $y^a$ are the transverse scalars. We remark that by adopting
an anzatz in which we set $y^5=0$ we may set the pull-back of the
background 3-form to the M5-brane worldvolume to zero so the
effective worldvolume action is indeed (\ref{e8}). We therefore
consider solitons with 2,3 or 4 active transverse scalars.

The
definition of the calibration forms proceeds in exactly the same
manner as for the flat computations with the co-ordinate basis
replaced by an orthonormal basis defined with respect to
${\tilde{g}}$. Thus for example the $SU(4)$ Kahler calibration
generalizes to a $SU(4)$ Hermitian calibration again
preserving $1 \over 16$ of the supersymmetry. The calibration form
is
\beq \f= {r^3 \over R^3} dx^1 \wedge dx^2 \wedge dx^3 \wedge dx^4
+dx^1 \wedge dx^2 \wedge dy^1 \wedge dy^2 + dx^3 \wedge dx^4
\wedge dy^1 \wedge dy^2.
\label{e10}
\eeq
It is required that $X^1+iX^2$ and $X^3+iX^4$ should
be holomorphic functions of $x^1+ix^2$ and $x^3+ix^4$.
More interesting examples may be obtained by considering
generalizations of special Lagrangian and exceptional geometries
\cite{G1}.
In all cases it is straightforward to verify that $d \f =
{\tilde{F}}$ so that the equations of motion are satisfied. All of
the examples in this background are supersymmetric, i.e.
\beq
\Gamma \ep = \ep
\label{e11}
\eeq
where for a p-brane with vanishing Born-Infeld fields
\beq
\Gamma = {1 \over (p+1)!} \ep^{\m_1 \dots \m_{p+1}} \gamma_{\m_1} \dots
\gamma_{\m_{p+1}}
\label{e12}
\eeq
and $\ep$ is a Killing spinor \cite{BOP}. It has been shown that just as
for
the flat background the calibration form may be constructed from
these Killing spinors satisfying appropriate constraints
\cite{G2}. Moreover the relation $d \f =
{\tilde{F}}$ may be seen to arise as a consequence of
the supersymmetry algebra.

The methods outlined here may be applied to p-brane configurations
in a large number of backgrounds for which the Born-Infeld type
fields vanish. In addition, an extension of this treatment has
been presented in \cite{E1}  which includes these fields.

\begin{acknowledgments}
I thank EPSRC for a studentship and the organizers for an excellent school.
\end{acknowledgments}



%

\end{document}